\documentclass[]{mn2e}

\usepackage{graphicx}
\usepackage[dvips]{color}

\newcommand{\kms}{\hbox{km~s$^{-1}$}}

\title[SIZE SCATTER OF PEGS AT $z\sim 2$]
{Large Size Scatter of Passively Evolving Lensed Galaxies at $z\sim 2$ in CLASH}
\author[Lulu Fan et al.]{Lulu Fan$^{1,2}$\thanks{E-mail:llfan@ustc.edu.cn},
Yang Chen$^{3,1}$,
Xinzhong Er$^4$,
Jinrong Li$^{1,2}$,
Lin Lin$^{1,2}$,Xu Kong$^{1,2}$ \\
$^1$Center for Astrophysics, University of Science and Technology of China, 230026 Hefei, China\\
$^2$Key Lab. for Research in Galaxies and Cosmology,USTC,CAS,230026,Hefei,China\\
$^3$Astrophysics Sector, SISSA, Via Bonomea 265, 34136 Trieste, Italy\\
$^4$National Astronomical Observatories, CAS, 100012 Beijing, China\\
}

\begin{document}

\date{}

\pagerange{\pageref{firstpage}--\pageref{lastpage}} \pubyear{0000}

\maketitle

\label{firstpage}

\begin{abstract}
In a systematic search over 11 cluster fields from Cluster Lensing And
Supernova survey with Hubble  (CLASH) we identify
ten passively evolving massive galaxies at redshift $z\sim2$. We
derive the stellar properties of these galaxies using HST WFC3/ACS
multiband data, together with Spitzer IRAC observations. We also
deduce the optical rest-frame effective radius of these high redshift
objects. The derived stellar masses and measured effective radii have
been corrected by the lensing magnification factors, which are estimated
by simply adopting the spherical NFW model for the foreground cluster
lens. The observed near-IR images, obtained by HST WFC3 camera with
high spatial resolution and lensed by the foreground
clusters, enable us to study the structures of such systems. Nine out of
ten galaxies have on average three times smaller effective radius than
local ETGs of similar stellar masses, in
agreement with previous works at redshift $1.5<z<2.5$. Combined with
literature data for $z\sim 2$, we find that the mass-normalized
effective radius scales with redshift as $r_e/M_\star^{0.56} \propto
  (1+z)^{-1.13}$. We confirm that their size distribution shows a
large scatter: from normal size to $\sim 5$ times smaller
compared to local ETGs with similar stellar masses. The 1-$\sigma$
scatter $\sigma_{{\rm log} r_e}$ of the size distribution is 0.22 and 0.34
at $z\sim 1.6$ and $z\sim 2.1$,respectively.The observed large size scatter
has to be carefully taken into account in galaxy evolution model predictions.
\end{abstract}


\section{Introduction}

Local Early-Type Galaxies (ETGs) are spheroidal systems,
which predominantly consist of old stars with mass-weighted
ages of $\geq 8-9$ Gyr  (see Renzini 2006 for a review). This
provides strong evidence that most of the stars in ETGs
were formed at redshift $z \geq 1.5$ and evolved passively. Recent
observations (e.g. Kriek et al. 2008) found red, massive ETGs already at $z > 2$. 
Many of these ETGs are designated as passively evolving galaxies (PEGs), generally
by a color-color selection criterion and/or a low specific star formation
rate (sSFR). Massive PEGs contribute a significant fraction
to massive galaxies at $z > 2$, with a number density of $\sim1\times 10^{-4} Mpc^{-3}$
(Kriek et al. 2008).

Sizes of local ETGs have been found to tightly correlate with stellar
masses (Shen et al. 2003). While compared to
local ETGs of similar stellar masses, the sizes of
massive ($M_\star\ge 10^{10.5}M_\odot$), PEGs at redshift $z\sim 2$ have
been found to be much smaller, by a factor of $\sim 3-4$ (Daddi et
al. 2005; Trujillo et al. 2006; Toft et al. 2007; van der Wel et
al.2008; van Dokkum et al. 2008, 2009; Damjanov et al. 2009; Saracco et
al. 2009; Cassata et al. 2010, 2011; Ryan et al. 2012; Newman et
al. 2012; Szomoru et al. 2012; Zirm et al. 2012). Deep HST/WFC3
images reveal that PEGs at $z\sim 2$ are not surrounded by the faint
extended envelopes of material (e.g. Szomoru et al. 2012) .
On the other hand, the lack of compact massive galaxies at
low redshift implies significant size evolution from $z\sim2$ to
$z\sim0$  (Trujillo et al. 2009).

Two popular models for the size evolution have been proposed:  (1)
dissipationless  (dry) minor merger  (Naab et al. 2009, Hopkins et
al. 2010);  (2) "puff-up" due to the gas mass loss by AGN  (Fan et
al. 2008, 2010) or supernova feedback  (Damjanov et al. 2009). However,
recent studies found that observed size evolution of the
compact massive galaxies at redshift $z\ge 1.5$ can not be explained
well by the present models.
Nipoti et al (2012) constructed a $\Lambda$CDM-based
analytic framework, supported by suites of N-body
merger simulations, to predict the evolution of ETGs
undergoing dry mergers. Using a compilation of
observations of ETGs at $z = 1-2.5$,
they concluded that mergers alone are not consistent
with the observed rate of structural evolution at $z\ge 1.5$.
For the puff-up model, Ragone-Figueroa \& Granato (2011)
found the mass loss by AGN feedback can produce a significant size increase
based on the results of numerical simulations. However, they augued most of the puffing up 
occurs when the stellar population are much younger than 
the estimated ages of compact high-z PEGs, so that  the puffing up due to the gas mass loss 
may not dominate the size evolution observed so far.

Size evolution with redshift has been described by $r_e\propto
 (1+z)^\beta$ in the literature (e.g. Newman et al. 2010, Damjanov et al. 2011, Cimatti et al. 2012).
However, the value of $\beta$ is diverse for different works,
varying from $\sim -0.8$ to $\sim -1.6$.  The large diversity
of $\beta$ may be due to the different sample selection criteria (morphological,
color-color or sSFR selections) and/or partly due to the sparsity
of data points at high redshift. For instance, there are only about ten objects at $z>2$
in Damjanov et al. 2012. Using $UVJ$ and sSFR selection criteria,
 Szomoru et al. (2012) selected 21 PEGs at $z\sim 2$ in the GOODS-South
field as part of the CANDELS survey. They found those 21 PEGs at
$z\sim 2$ not only had smaller effective radii than local ETGs of similar stellar masses,
but also had a significant spread in their size distribution  ($\sigma_{{\rm log} r_e} \sim 0.24$).
If the larger size scatter is real at higher redshift, it will be another observational constraint
for dissecting the size evolution models.

In this letter,we will use multiband data from the CLASH
survey to search massive PEGs at $z\sim2$. HST/WFC3 F160W-band images will be used to
derive the sizes of PEGs. And complementary with the recent archive data,
we will study size distribution at $z\sim 2$ and size evolution with redshift. Throughout this
letter, we assume a concordance $\Lambda$CDM cosmology with
$\Omega_{\rm m}=0.3$, $\Omega_{\rm \Lambda}=0.7$, $H_{\rm 0}=70$ \kms
Mpc$^{-1}$. All magnitudes are in the AB systems.

\section{Data and Sample Selection}

The Cluster Lensing And Supernova survey with Hubble  (CLASH) is an
\emph{HST} Multi-Cycle Treasury program that is acquiring images in
16 broad bands with the Wide Field Camera 3  (WFC3) UVIS and IR
cameras, and the Advanced Camera for Surveys  (ACS) for 25
clusters (Postman et al. 2012). The images are reduced and mosaiced  ($0."065$ pixel$^{-1}$)
using the \emph{MosaicDrizzle} pipeline  (Koekemoer et
al.2011). Photometric redshifts are estimated for all the galaxies
using the full 16-band photometry via both the \emph{BPZ}  (Ben\'{i}tez
2000) and \emph{LePhare}  (Arnouts et al. 1999) codes. In this letter,
we use the released photometric catalog of 11 clusters
\footnote{http://archive.stsci.edu/prepds/clash/}. We supplement the
\emph{HST} observations with \emph{Spitzer} IRAC data. We use a
software package with object template-fitting method  (TFIT; Laidler et
al. 2007) to measure IRAC photometry. TFIT uses the spatial positions and morphologies of objects
in a high-resolution image to construct templates, which are then fitted to a lower resolution
image. The fit can be simultaneously done for several close objects in a crowded field.
We use the F160W band as  the TFIT high-resolution template to measure the IRAC photometry.

Stellar masses and star formation rates  (SFRs) are estimated from SED
fitting to the full photometric data set  (see Figure 1), assuming a
Kroupa IMF  (Kroupa 2001) and the BC03 stellar population models
 (Bruzual \& Charlot 2003). We use a Bayesian-like code CIGALE
\footnote{http://cigale.oamp.fr/}  (Noll et al. 2009) to carry out the
SED fitting. As described in Noll et al. 2009 , an exponentially decreasing star formation
history and a modified Calzetti attenuation law  have been adopted
(see Calzetti et al. 2000 for more details).
The SED-derived stellar masses and SFRs are corrected by
the foreground lens magnification factor. We simply adopt the
spherical NFW  (Navarro et al. 1997) model for the foreground cluster
lens.  The lensing properties of NFW halo can be calculated
analytically for given position  (Bartelmann 1996). The mass and
concentration of the lens clusters are obtained from the literature
 (Umetsu et al. 2012, Zitrin et al. 2011, Gavazzi et al. 2003 and  Kling et al. 2005).
The detailed properties of the lens clusters are not taken into account, such as mass ellipticity,
substructures, etc. The estimated magnifications may be slightly
underestimated, especially for the clusters not long after merging.

\begin{figure}
\begin{center}
\includegraphics[width=1.0\linewidth]{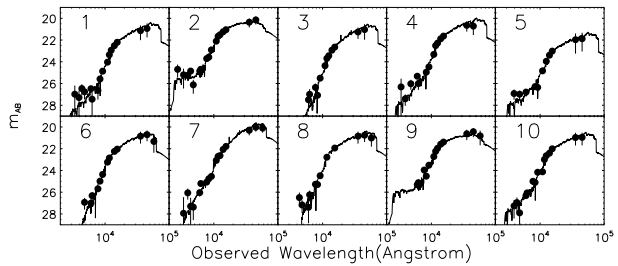}
\caption{Broad band SEDs  (black points) and best-fit models (solid
  lines) of $z\sim2$ PEGs. On the top-left of each panel, the number
  marks the galaxy ID in our sample (see Table 1). }
\label{test figure}
\end{center}
\end{figure}

We select passively evolving galaxies using a selection criterion
based on sSFR, with sSFR$<
0.3/t_H$, where $t_H$ is the Hubble time. As additional selection criteria, 
we also demand that the PEGs will be at  $1.5<z<2.5$ and stellar masses $M_\star >
10^{10.5}M_\odot$. The final sample consists of ten galaxies.
Our selection criterion is very close to those using
a fixed sSFR (e.g. $sSFR<10^{-11} yr^{-1}$ ). And we also
check if our selection criterion is consistent with the
color-color selection. We find eight out of ten PEGs
in our sample fulfill the $J-3.6 \mu m$ and $z-J$ color-color
selection (Papovich et al. 2012).  $J-3.6\mu m$ and $z-J$ colors
correspond approximately to rest-frame $V-J$ vs. $U-V$ at $z\sim 2$,
which Williams et al. (2009) showed can effectively separate
passively evolving galaxies from star-forming galaxies.

\begin{table*}
\begin{center}
\caption{The Sample of Passively Evolving Lensed Galaxies at $1.5<z<2.5$}
\begin{tabular}{l l c r c l l l c l l c}
\hline
Cluster$^{ (a)}$       &  ID$^{ (b)}$ & Ra$^{ (c)}$       & Dec$^{ (c)}$   & $\theta ^{ (d)}$     &$\mu ^{ (e)}$   & $z^{ (f)}$  &  {\rm log} $M_\star ^{ (g)}$  & {\rm log} sSFR$^{ (h)}$    &  n$^{ (i)}$  &  $r_e^{ (j)}$ &  b/a$^{ (k)}$ \\
              &     & [deg]     & [deg]     &  [arcsec]    &     &              &  [$M_\odot$]           & [yr$^{-1}$] &             &  [kpc] \\
\hline
MACSJ1206-08  &   1 & 181.53618	        & -8.81180  & 64.7  & 1.83 & 1.998 & 10.85 & -11.15 &  6.3    & 1.4          & 0.39 \\
MACSJ0647+70  &   2 & 101.99016	& 70.26104  & 57.4  & 1.13 & 1.556 & 11.08 & -10.88 &  3.9    & 5.1  & 0.75 \\
MACSJ0647+70  &   3 & 101.99335	& 70.23792  & 55.6  & 1.15 & 1.941 & 10.99 & -11.23 &  3.1    & 0.9  & 0.50 \\
MACSJ0647+70  &   4 & 101.99383	& 70.23749  & 57.3  & 1.15 & 1.942  & 11.20 & -11.17 &  6.0    & 1.5  & 0.53 \\
MACSJ0744+39  &   5 & 116.21017	& 39.43837  & 74.0  & 1.14 & 2.276 & 10.88 & -10.91 &  1.8    & 0.7  & 0.62 \\
MS 2137-2353  &   6 & 325.07814	       & -23.67539 & 71.2  & 1.38 & 1.534 & 10.74 & -11.16 &  2.2    & 1.0         &  0.36 \\
RXJ1347-1145  &   7 & 206.90139	       & -11.74505 & 84.9  & 2.03 & 1.646  & 10.77 & -10.76 &  2.0    & 2.2         & 0.64 \\
RXJ1347-1145  &   8 & 206.86622        & -11.76696 & 64.1  & 3.30 & 1.811 & 10.56 & -11.27 &  4.5    & 2.0        &  0.32 \\
RXJ1347-1145  &   9 & 206.88078	       & -11.77017 & 58.3  & 3.84 & 1.662 & 10.54 & -10.96 &  3.0    & 1.3       & 0.47  \\
MACSJ1149+22  &   10& 177.40348      & 22.41853  & 74.4  & 1.09  & 2.007  & 11.04 & -11.13 &  4.0$^{ (l)}$& 0.9  & 0.87 \\
\hline
\end{tabular}
\parbox{180mm} { (a): The foreground cluster; (b): ID of PEGs;
 (c): Right ascension and Declination of PEGs; (d): Projected angular
 distance of PEG from the cluster center; 
 (e): The lensing magnification factor which is estimated by simply 
 adopting the spherical NFW profile for the foreground cluster
 lens. The mass and concentration of the lens clusters are obtained 
 from literature  (Umetsu et al. 2012, Zitrin et al. 2011, 
 Gavazzi et al. 2003 and  Kling et al. 2005);  
 (f): Photometric  redshift of PEGs using BPZ ; (g): Stellar masses , which
 have been homogenized to BC03 models 
 and Kroupa IMF ;  (h): Specific star formation rate ;
 (i): Sersic index ;  (j) : Effective radius, which has been corrected
 by the lensing magnification factor ;  (k) : Axis ratio ;  (l) : Sersic
 index is fixed to 4.0 .
}
\end{center}
\end{table*}

\section{Sizes of PEGs at $z\sim2$}

\begin{figure}
\includegraphics[width=1.\linewidth]{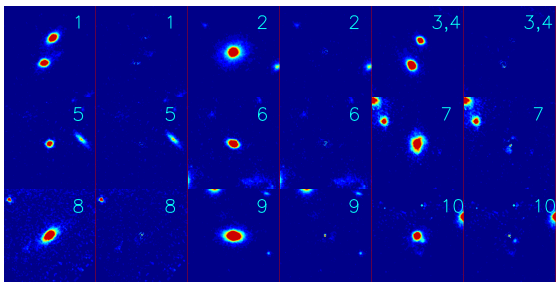}
\caption{HST F160W band original (left) and residual (right) images of
  passively evolving galaxies in our sample. The size of each panel is
  $6\times6$ arcsec$^2$. On the top-right of each panel, the number
  marks the galaxy ID in our sample.  }
\end{figure}

We use the \emph{GALFIT} package  (Peng et al. 2002) to fit S\'{e}rsic
profiles to the PEGs in the HST F160W band. The fits are performed in a $6\times6$
arcsec$^2$ region (See Figure 2).  Empirical PSFs instead of model
PSFs are used in the fits. We obtain the empirical PSF for each
cluster using PSFEx  (Bertin 2011). PEG-1 is close to another
galaxy. Two independent S\'{e}rsic models are simultaneously fitted to
PEG-1 and its companion, respectively. We use the same procedure to
fit PEG-3 and PEG-4  (both of them are PEGs). The S\'{e}rsic index n is
fixed to 4 for PEG-10 in order to make the fit converge.

The observed images of high redshift PEGs in our sample have been
lensed by the foreground clusters. We need to take the lens
magnification into account in order to estimate the intrinsic sizes of
PEGs. For simplicity, we assume that the lensed image has
the same light profile as the source. Following the definition of the
magnification factor $\mu$, we have
\begin{equation}
r_{s}=\frac{r_{d}}{\sqrt{\mu}},
\end{equation}
where $r_{s}$ is the intrinsic angular size of the source image, $r_{d}$ is
the measured angular size of the lensed image,both in unit of arcsec.
$\mu$ is the lensing magnification
factor estimated using the method introduced in the Section 2 
(see Bartelmann \& Schneider 2001 for more detail on lensing). The
measured angular size from the best-fit S\'{e}rsic model is
corrected using the formula in Equation 1 and then converted to physical size
using photometric redshift.

\section{Discussion}

\begin{figure}
\includegraphics[width=1.0\linewidth]{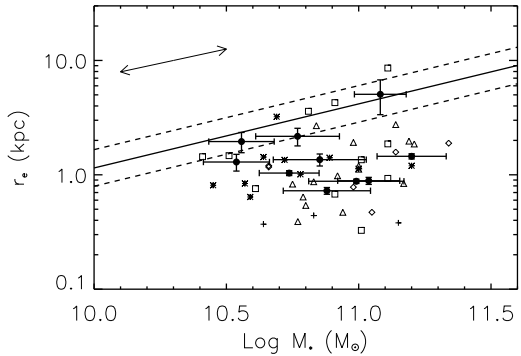}
\caption{The distribution of effective radius versus stellar mass for
  passively evolving galaxies at $1.5<z<2.5$. Our PEGs sample is
  represented by black filled circles. Both stellar masses and
  effective radius are corrected by the lensing magnification factor. The
  $\pm 1 \sigma$ errors of stellar masses and effective radius are
  given by CIGALE and GALFIT, respectively. Other literature data are over-plotted
  with plus signs  (Cassata et al. 2010), asterisks  (Cimatti et
  al. 2008), diamonds  (Zirm et al. 2012), triangles  (Szomoru et
  al. 2012) and squares  (Ryan et al. 2012), respectively. All of the
  stellar masses are homogenized to BC03 and Kroupa IMF.
   The solid line shows the local
  effective radius and stellar mass relation of early-type
  galaxies (Shen et al. 2003). The dashed lines indicate the $\pm 1
  \sigma$ scatter of Shen et al. (2003) relation. The arrows on
  top-left show the moving directions due to the uncertainty of
  the lensing magnification factors $\mu$.  }
\end{figure}

van der Wel et al. (2011) argued the majority of PEGs at $z\sim 2$ in
their sample were disk-dominated. We investigate this argument with
our sample. Half of our PEGs have best-fitting axis ratios $b/a \leq
0.5$ and six  (PEG-1, PEG-3, PEG-4, PEG-6, PEG-8 and PEG-9) have $b/a \leq 0.6 $  (See Table
1). The axis ratio distribution indicates that a significant fraction
of PEGs at $z\sim 2$ may be disk-dominated. However, we find that 
S\'{e}rsic index $n$ are mostly greater than 2, indicating
elliptical-like morphology.  Only two galaxies  (PEG-5 and PEG-7) have
$n\leq 2$. We re-fit our PEGs with two S\'{e}rsic models. We fix the
index $n$ of one S\'{e}rsic model to 4 to represent the bulge
component and $n=1$ to represent the disk component. Two galaxies
 (PEG-7 and PEG-9) show a significant disk component with $\geq 50\%$
light in disk. The results by axis ratios and two-component model show that $\sim 20-60\%$
of PEGs at $z\sim 2$ are disk-dominated in our sample.

In Figure 3, we plot the distribution of effective radius versus
stellar mass for our PEGs sample at $1.5<z<2.5$. In our sample, both
stellar mass and effective radius are corrected by the lensing
magnification factor. We also compare with the local relation  (Shen et
al. 2003) and other literature results with similar range of redshift
and stellar masses  (Cimatti et al. 2008, Cassata et al. 2010, Szomoru
et al. 2012, Ryan et al. 2012 and Zirm et al. 2012). All stellar masses
are homogenized to BC03 and Kroupa IMF. We find that our results are
consistent with literature results at $z\sim 2$. Nine out of ten PEGs
in our sample are below the local relation, with $r_e \sim 1$ kpc.
These galaxies are on average $\sim 3-4$ times smaller than the local
counterparts with similar stellar masses. The compact sizes of high
redshift PEGs can be formed naturally according to the dissipative
collapse of baryons  (see Fan et al. 2010).

In Figure 3, we find that the sizes of PEGs at $z\sim 2$ span a wide
range and need a significant evolution to catch up the local
relation. Furthermore, we explore the size evolution with redshift
combining our sample together with the literature data. We divide the combined data into two redshift
bins, $z\sim 1.6$ and $z\sim 2.1$.  The former redshift bin includes
25 PEGs and the later bin has 31. The average stellar masses of both
bins are $\sim 8\times 10^{10} M_\odot$. We adopt the local size-mass
relation $r_e \propto M_\star^\alpha$, where $\alpha$ equals 0.56
by Shen et al.  (2003). We plot the average mass-normalized sizes of
two redshift bins in Figure 4  (filled squares). We also plot the local
relation with shaded area in Figure 4.  We use the relation
$r_e\propto  (1+z)^\beta$ to describe the redshift evolution. For the
combined data, we have $\beta \sim -1.13\pm 0.13$  (See the dashed line in
Figure 4). The $\beta$ value is in agreement with the result of
Cimatti et al. (2012). They derived $\beta \sim -1.06\pm 0.14 $ for the
mass-normalized radius with ${\rm log} M_\star/M_\odot >
10.5$. However, other results with steeper slope had been found from
different samples. Using optical and near-infrared photometry in the
UKIRT Ultra Deep Survey and GOODS-South fields of the CANDELS survey,
Newman et al. (2012) found the size evolution of quiescent galaxies
with redshift could be described as ${\rm log}
 (r_e/M_\star^{0.57})=0.38-0.26 (z-1)$.  This relation can be
approximately rewritten as $r_e/M_\star^{0.57}\propto  (1+z)^\beta$,
with $\beta\sim -1.33$. Damjanov et al. (2011) found an even steeper
value, $\beta \sim -1.62\pm 0.34$.  The sizes need to increase by a factor of
$\sim 3$ for $\beta \sim -1.0$ from $z= 2$ to 0.  While for
$\beta \sim -1.6$, the sizes would increase by a factor of $\sim 6$
during the same comic epoch. The diversity of observed size evolution
with redshift is at least partly due to the sparsity of data points at
$z\sim 2$. A large number sample of PEGs at $z\sim 2$ will be helpful
to deicide the valid size evolution with redshift and constrain the
different galaxy evolution models.

\begin{figure}
\includegraphics[width=1.0\linewidth]{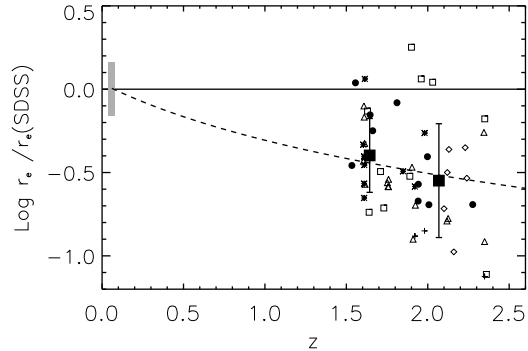}
\caption{Evolution of the effective radius with redshift. The shaded
  area reflects the distribution of local SDSS galaxies.  The filled
  squares show the average sizes of PEGs at $z\sim 1.6$ and $z\sim
  2.1$.  The error bars show the $1 \sigma$ scatter. The dashed line
  represents the size evolution with redshift described by $r_e\propto
   (1+z)^{-1.13}$. The other symbols are the same as those in Figure 3.  }
\end{figure}

A relevant feature of the data for massive PEGs at high redshift
is the quite large scatter of the size, as is apparent from Figures 3
and 4.  We plot the $1\sigma$ size scatter  ($\sigma_{{\rm log}{r_e}}$)
of two redshift bins in Figure 4.  We find that the size scatter is
large at $1.5<z<2.5$ in the combined data.  The
values of $\sigma_{{\rm log}{r_e}}$ are 0.22 and 0.34 at $z\sim 1.6$
and $z\sim 2.1$, respectively. Newman et al. (2012) showed a very similar
result:$\sigma_{{\rm log}{r_e}}=$ 0.24 and 0.26 at $1.0<z<1.5$ and $1.5<z<2.0$,
respectively.In Figure 4, the error bars show 1$\sigma$
size scatter of two redshift bins and the shaded area reflects the
distribution of local SDSS galaxies.  Compared to the small size
scatter of SDSS ETGs  ($\sigma_{{\rm log}{r_e}}= 0.16$ at $z\sim 0.06$),
the size scatter also needs a significant evolution with redshift
 (about a factor of $1.5-2$ from $z\sim 2$).

Larger size scatter at higher redshift need to be accounted for by
galaxy evolution models.  If the observed size evolution with redshift
is dominated by the dry minor merger, we will expect that more compact
PEGs have higher dry merger rates relative to less compact objects at
given redshift range. Another possibility is size evolution at $z\sim
2$ could be dominated by AGN feedback. One assumption is that gas
will be removed by the strong AGN feedback in a short timescale. In
this case, galaxy structure after AGN feedback will relax to the new
equilibrium.  Another situation for AGN feedback mechanism is that the
gas removal will not complete in only one single shot. Martizzi et
al. (2012) used cosmological simulations to show the effect of AGN
feedback on the gas distribution in the central regions of cluster
galaxies. They found gas heated and expelled by AGN feedback can
return after cooling; repeated cycles could modify the stars and dark
matter mass profile. During the cycles, the observed sizes will
oscillate from compact  (all gas is located within the central regions of
galaxies) to normal  (gas is expelled from the galaxy center).  This
produces the large size scatter observed at high redshift, and the
size scatter will be related to AGN activities which decrease from
$z\sim 2$ to $z\sim 0$.  For AGN feedback mechanism, larger gas
fraction at higher redshift will also be needed in order to explain
the observed size scatter evolution. Most recently, Olsen et al. (2012) found
the ubiquitous presence of AGN in massive, quiescent $z\sim 2$ galaxies.
This result provided observational indications for the important
role AGN played in explaining the size  (and size scatter) evolution.

It should be noticed that the lensing magnification correction that we adopted is simple,
which may introduce uncertainties as well.  However, the uncertainty on
the lensing magnification may not significantly alter our main results above.
The corrected stellar mass is proportional to $1/{\mu}$, while
the corrected effective radius is proportional to $1/\sqrt{\mu}$.
As $\mu$ decreases/increases, the location on the size-mass plane will move along the directions
shown by arrows in Figure 3.  The arrows  ($r_e\propto M_\star^{0.5}$)
nearly parallel to the local size-mass relation  ($r_e\propto M_\star^{0.56}$).
So the uncertainty of $\mu$ will have a small effect on the mass-normalized effective radius.
For instance, the mass-normalized effective radius with $M_\star \sim 10^{11}M_\odot$
will only change by a factor of $<5\%$
when $\mu$ increases by a factor of two.

\section{Summary}

In this letter, we use the CLASH survey data to select a massive, lensed PEGs
sample at $1.5<z<2.5$, based on the SED-fitting sSFR and accurate
photometric redshift.  We deduce the optical rest-frame effective radius of
these high redshift objects using GALFIT.  Combined with literature
data, we find the size evolution of passively evolving galaxies scales
with redshift as $r_e\propto  (1+z)^{-1.13\pm 0.13}$.  We confirm that their
sizes have a large spread.  The 1-$\sigma$ scatter $\sigma_{{\rm log}
  r_e}$ of the size distribution is 0.22 and 0.34 at $z\sim 1.6$ and
$z\sim 2.1$, which is about 1.4  and 2.1  times larger than the local
value ($\sigma_{{\rm log} r_e}\sim 0.16$), respectively. The observed large size scatter
has to be carefully taken into account in galaxy evolution model predictions.

\section*{Acknowledgements}

 We thank the  anonymous referee for the careful reading and the valuable comments that helped improving our paper.
 We thank Prof. Wei Zheng for discussion on CLASH data. We also thank Prof Xianzhong Zheng and Dr. Xinwen Shu for valuable
suggestions on Spitzer IRAC data analysis.
This work has been supported by the Chinese National Science Foundation  (NSFC-11203023) and Chinese Universities Scientific Fund  (WK2030220004).
L.F. thanks the partly financial support from the China Postdoctoral Science Foundation (Grant No.:2012M511411).
We specially thank the CLASH team making the catalogues and images available for public.
This work is also partly based on archival data obtained with the Spitzer Space Telescope.

\label{lastpage}
\end{document}